\documentclass[conference]{IEEEtran}
\IEEEoverridecommandlockouts
\usepackage{cite}
\usepackage{amsmath,amssymb,amsfonts}
\usepackage{algorithmic}
\usepackage{graphicx}
\usepackage{textcomp}
\usepackage{xcolor}
\usepackage{hyperref}
\hypersetup{colorlinks,linkcolor={blue},citecolor={blue},urlcolor={blue}}

\def\BibTeX{{\rm B\kern-.05em{\sc i\kern-.025em b}\kern-.08em
    T\kern-.1667em\lower.7ex\hbox{E}\kern-.125emX}}

\begin{document}

\title{Learning Exemplar Representations \\ in Single-Trial EEG Category Decoding
\thanks{This publication emanated from research supported in part by Science Foundation Ireland research grants 16/RI/3399, 18/CRT/6049, and 20/FFP-P/8853, the United States National Science Foundation Grant 1734938-IIS, and the Defense Advance Research Projects Agency. This publication is approved for public release; distribution is unlimited. Any opinions, findings, views, conclusions or recommendations expressed in this material are those of the authors and do not necessarily reflect the views, official policies, or endorsements, either expressed or implied, of the sponsors. Finally, we thank Hari M. Bharadwaj for suggesting that we look into the issue discussed in this paper as a potential source of leakage.
}

}

\author{\IEEEauthorblockN{Jack A. Kilgallen}
\IEEEauthorblockA{\textit{Hamilton Institute} \\
\textit{Maynooth University}\\
Maynooth, Ireland \\
\href{mailto:jkilgallen@protonmail.com}{jkilgallen@protonmail.com}}
\and
\IEEEauthorblockN{Barak A. Pearlmutter}
\IEEEauthorblockA{\textit{Department of Computer Science} \\
\textit{Maynooth University}\\
Maynooth, Ireland \\
\href{mailto:barak@pearlmutter.net}{barak@pearlmutter.net}}
\and
\IEEEauthorblockN{Jeffrey Mark Siskind}
\IEEEauthorblockA{\textit{Elmore Family School of Electrical} \\
\textit{and Computer Engineering} \\
\textit{Purdue University}\\
West Lafayette, IN, USA \\
\href{mailto:qobi@qobi.org}{qobi@qobi.org}}
}

\maketitle

\begin{abstract}
    Within neuroimgaing studies it is a common practice to perform repetitions of trials in an experiment when working with a noisy class of data acquisition system, such as  electroencephalography (EEG) or magnetoencephalography (MEG). While this approach can be useful in some experimental designs, it presents significant limitations for certain types of analyses, such as identifying the category of an object observed by a subject. In this study we demonstrate that when trials relating to a single object are allowed to appear in both the training and testing sets, almost any classification algorithm is capable of learning the representation of an object given only category labels. This ability to learn object representations is of particular significance as it suggests that the results of several published studies which predict the category of observed objects from EEG signals may be affected by a subtle form of leakage which has inflated their reported accuracies. We demonstrate the ability of both simple classification algorithms, and sophisticated deep learning models, to learn object representations given only category labels. We do this using two datasets; the Kaneshiro et al. (2015) dataset and the Gifford et al. (2022) dataset. Our results raise doubts about the true generalizability of several published models and suggests that the reported performance of these models may be significantly inflated.
\end{abstract}

\begin{IEEEkeywords}
EEG, machine learning, leakage, category decoding
\end{IEEEkeywords}

\section{Introduction}
Within neuroimgaing studies when working with a noisy modality such as electroencephalography (EEG), or magnetoencephalography (MEG), it is common practice to perform multiple trials using a single stimulus. With some experimental designs this repetition of trials can be useful for reducing the impact of noise on the analysis being performed, particularly when extracting features such as Event Related Potentials (ERPs)\cite{davis_effects_1939}. However, this practice presents significant limitations for certain types of analyses such as \emph{single-trial category decoding}.

In single-trial category decoding experiments, subjects are shown a series of objects from different categories while their EEG signals are recorded. The aim of the experiment is to identify the category of an object observed by a subject from a single-trial of their EEG signals. This is a challenging task as the EEG signals are noisy, and the features which are relevant to the category of an object may be subtle.

When performing single-trial category decoding, the use of multiple trials per object can allow a model to use information which we would not expect it to have access to at prediction time \emph{(leakage)}. When multiple trials per object are used, a classification algorithm may overfit to features of an EEG signal which are specific to an observed object from a category \emph{(an exemplar representation)} rather than to features of neural responses which are shared by many or all objects from a category \emph{(a category representation)}. Moreover, if trials relating to specific stimuli are allowed to appear in both the training and testing sets, then algorithms which overfit to exemplar representations are likely to achieve higher accuracies. This occurs as their performance is evaluated using trials which are more similar to the trials they were trained on than would be expected in a real-world scenario. In this way a loss of generalizability is rewarded by the evaluation procedure.

An analogy to this form of leakage can be found in the case of a machine learning student who wants to classify images of cats and dogs. The student collects photos from friends and family of their cats and their dogs, but each person they ask sends multiple photos of each of their pets. If the student trains a computer vision algorithm to classify the images of cats and dogs, but images relating to specific cats or dogs are allowed to appear in both the training and testing sets, then the algorithm may be affected by a similar form of leakage. Instead of learning what a dog or a cat generally looks like, the algorithm may learn what a specific dog or cat looks like. It would then achieve higher accuracy by overfitting to the features of that specific dog or cat (i.e. learning the representation of an exemplar). The situation would be further exacerbated if the dataset didn't contain many breeds of dogs or cats, as the algorithm would have fewer opportunities to learn what features are shared by all dogs or cats. Similarly, in the case of single-trial category decoding, algorithms may achieve increased by overfitting to features of a signal a subject's brain generates when they observe a specific object.

While this form of leakage may seem contrived, it is increasingly present in recent literature. Since the publication of the Kaneshiro et al.\ (2015)\cite{kaneshiro_representational_2015} dataset (hereafter referred to as the Kaneshiro dataset) which features multiple trials per stimulus, 12 studies have made use of their dataset for single-trial EEG category decoding\cite{ahmadieh_hybrid_2023, bobe_single-trial_2018, jiao_decoding_2019, bagchi_adequately_2021, bagchi_eeg-convtransformer_2022, deng_eeg-based_2023, fares_brain-media_2020, kalafatovich_decoding_2020, kalafatovich_learning_2023, kalafatovich_subject-independent_2021, luo_dual-branch_2023, yavandhasani_visual_2022}. However, none of these studies make any mention of the potential for leakage in their experimental design, and so it is likely that the results of these studies are affected.

To investigate the ability of classification algorithms to exploit exemplar representations to inflate their performance at single-trial category decoding, we augmented two EEG category decoding datasets to remove any semantic meaning from their category labels. We remove semantic meaning by creating new category labels for which no reasonable distinction can be made between categories. Therefore, when performing category decoding on such a dataset, any model which achieves a decoding accuracy which is above chance, in a statistically significant manner, would suggest that the model has exploited knowledge of an exemplar representation.

Our analysis demonstrates not only that some bespoke EEG decoding algorithms are capable of learning exemplar representations given only category labels, but that even simple classification algorithms such as k-Nearest Neighbors (k-NN) are capable of exploiting exemplar representations. Moreover, we demonstrate that this form of leakage can occur using two separate datasets; the Kaneshiro datset\cite{kaneshiro_representational_2015} and the Gifford et al.\ (2022)\cite{gifford_large_2022} dataset (hereafter referred to as the Gifford dataset). 

\section{Related Work}
While machine learning has a broad range of applications, training models on data without a strong understanding of the domain the data originates from can cause confounds and leakage to go unnoticed. This has occurred in a variety of contexts\cite{REP-Workshop-2022} and attention is being given to rooting out studies affected by such confounds\cite{kapoor_leakage_2023}. Even within a domain which might be considered niche, such as neuroimaging, significant confounds, such as the Block Design Confound\cite{li_perils_2021}, have been discovered in recent literature. The work presented here is part of a sustained effort to identify erroneous or exaggerated claims resulting from a misapplication of machine learning techniques in neuroimaging studies.

\section{Materials and Methods}
    \subsection{Datasets}
    The datasets used in this study are the Kaneshiro dataset\cite{kaneshiro_representational_2015} and the Gifford dataset\cite{gifford_large_2022}. They were selected as they were publically available, and possessed both multiple trials per stimulus, and a category hierarchy. 
    \subsubsection{The Kaneshiro Dataset\cite{kaneshiro_representational_2015}}
        This dataset consists of preprocessed EEG recordings taken from 10 subjects while they viewed 72 images evenly distributed across 6 categories: Human Body (HB), Human Face (HF), Animal Body (AB), Animal Face (AF), Fruit/Vegetable (FV) and Inanimate Object (IO). Fig.~ref{kaneshiro-category-structure} shows the stimuli that were presented to the subjects grouped by category. To reduce the impact that noise would have on their analysis the dataset contains 72 trials for each image per subject which were presented in random order. This gives a total of 5,184 trials per participant. The data was recorded using a 128 channel EEG system with a sampling rate of 1~kHz. The EEG signals were then preprocessed using a high-pass fourth-order Butterworth filter to attenuate frequencies below 1~Hz, and a low-pass Chebyshev Type I filter to attenuate frequencies above 25~Hz. Ocular artifacts were removed using the Bell-Sejnowski Infomax independent component analysis algorithm\cite{bell-sejnowski-1995}, and finally the data was subsampled to 62.5~Hz to reduce the computational cost of the analysis. 

        \begin{figure}[t]
            \centerline{\includegraphics[width=\columnwidth]{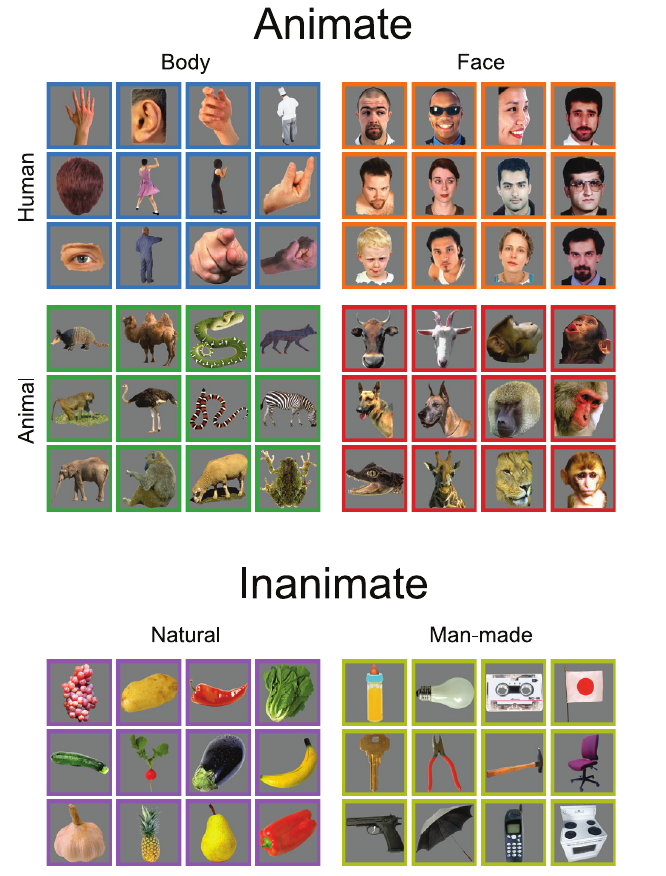}}
            \caption{The stimuli presented to subjects in the Kaneshiro dataset\cite{kaneshiro_representational_2015}.}
            \label{kaneshiro-category-structure}
            \end{figure}
        
        \subsubsection{The Gifford Dataset\cite{gifford_large_2022}}
        This dataset is made up of preprocessed EEG recordings taken from 10 subjects while they viewed a selection of images from the THINGS image database\cite{herbart_things_2019}. The images were divided across 27 high-level categories, which span 1854 object concepts. The object categories were partitioned into training and test sets with 1654 object concepts used in the training set and 200 used in the test set. For the training set 10 exemplars per concept which were each presented to the subjects once for a total of 16,540 trials per subject. For the test set its 200 object concepts each contained a single exemplar which was presented 80 times, for a total of 16,000 trials per subject. The experiment used a rapid serial visual presentation (RSVP) paradigm. During image presentation the subjects were asked to perform a target detection task in which they counted the number of times a predefined target image was presented. The data was recorded using a 64-channel EEG system sampled at 1000Hz, which was filtered using a 0.1Hz high-pass filter and a 100Hz low-pass filter, and baseline correction was applied using pre-stimulus recordings. The data finally subsampled to 100Hz, and 17 channels over the occipital and parietal cortex were selected. It should be noted that this dataset is not easily misused in the same way as the Kaneshiro dataset\cite{kaneshiro_representational_2015}, as it explicitly supplies the data as separate training and testing sets with no exemplars appearing in both sets. However, its large size and repetition of exemplars in its test set make it a very useful dataset to investigate if category decoding algorithms can learn exemplar representations when the dataset features a higher number of exemplars per category. 

        \subsection{Generating Pseudocategories}
        To demonstrate that category decoding algorithms can learn exemplar representations, we constructed \emph{pseudocategories} from the stimulus sets used in the original datasets. By pseudocategory we mean a collection of exemplars with no semantic relationship to each other, but which are presented to the model as if they were from the same category. Importantly, we ensured that the pseudocategories were balanced in both the number of exemplars they contained, and the distribution of those exemplars among their true categories.
    
            \subsubsection{The Kaneshiro Dataset\cite{kaneshiro_representational_2015} Pseudocategories}
            To construct the pseudocategories for this dataset, we selected a single exemplar from each of the 6 categories and combined all trials relating to these exemplars into a single pseudocategory. This was repeated 12 times until all exemplars were used, resulting in 12 pseudocategories. As there are 12 pseudocategories, which are balanced in their number of trials, it follows chance accuracy for a model predicting the pseudocategory of a trial is $\frac{1}{8}=8.25\%$.
            % Fig.~\ref{fig:kaneshiro-pseudocategories} shows the pseudocategories constructed.
    
            \begin{figure}[t]
                \centerline{\includegraphics[width=\columnwidth]{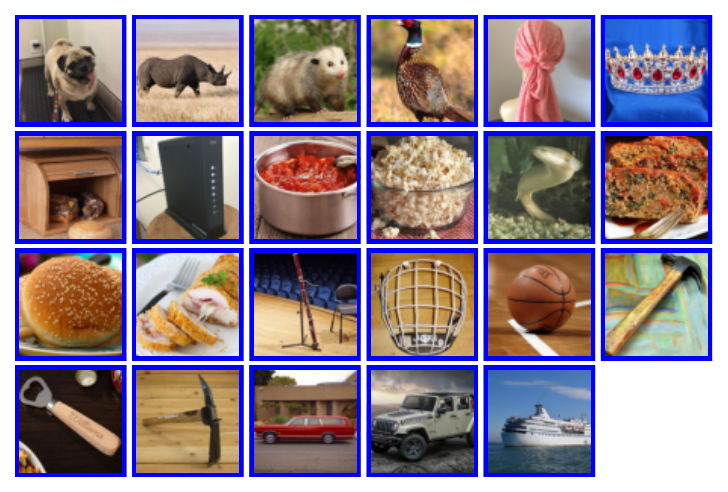}}
                \centerline{\includegraphics[width=\columnwidth]{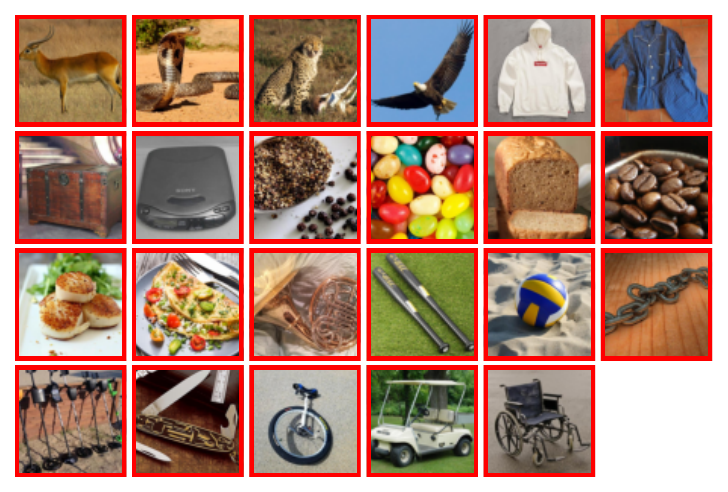}}
                \caption{Two of the pseudocategories generated for the Gifford dataset\cite{gifford_large_2022}.}
                \label{fig:gifford-pseudocategories}
                \end{figure}
    
            \subsubsection{The Gifford Dataset\cite{gifford_large_2022} Pseudocategories}
            To construct the pseudocategories for this dataset, we used only trials from the predefined test set for each subject as only this set contained multiple trials per exemplar. We then grouped them by their higher-level categories, and excluded any categories which contained fewer than five exemplars, or were deemed to be too similar to other high-level categories. Finally, we evenly divided exemplars from each category into 5 pseudocategories. The pseudocategories each contained: 3 animals, 1 bird, 2 items of clothing, 1 container, 1 electronic device, 6 items of food, 1 musical instrument, 2 items of sports equipment, 3 tools, and 3 vehicles for a total of 23 exemplars per pseudocategory. 
            Fig.~\ref{fig:gifford-pseudocategories} shows an example of a pseudocategory constructed this way. Again, as the pseudocategories are balanced in their number of trials, and there are 5 categories, it follows that chance accuracy for a model predicting the pseudocategories of a trial is $\frac{1}{5}=20\%$.
        
    \subsection{Category Decoding Algorithms}
    To test the ability of classification algorithms to learn exemplar representations, we selected a broad variety of models. Our selection was based on capturing the spectrum of complexity of classification algorithms commonly used in EEG decoding, from simple classical machine learning algorithms to state-of-the-art deep learning models. While many recent models have been developed specifically for single-trial EEG category decoding the reproducibility of these models is limited by the availability of their implementations. Therefore, we selected models which are well known and have implementations available in popular libraries. The models used in this study are:
    \begin{itemize}
        \item Support Vector Machine (SVM)\cite{cortes_support_1995}.
        \item Linear Discriminant Analysis (LDA)\cite{fisher_use_1936}.
        \item EEGNet\cite{lawhern_EEGNet_2018}.
        \item DeepConvNet\cite{schirrmeister_deep_2017}.
        \item ShallowConvNet\cite{schirrmeister_deep_2017}.
    \end{itemize}
    The k-Nearest Neighbors (k-NN) algorithm\cite{cover_nearest_1967} was selected to investigate the ability of a simple classical machine learning algorithm to learn exemplar representations.
    The neural network models were implemented using the PyTorch library\cite{paszke_pytorch_2019} with some modifications to the original implementations to account for the lower temporal resolution of our data relative to the dataset these models were originally trained on. The scikit-learn library\cite{pedregosa_scikit-learn_2011} was used to implement the SVM, LDA and kNN models. A detailed description of the model architectures and training procedures can be found in the supplementary material.

\section{Learning Exemplar Representations from Categories}

        \subsection{Model Implementations and Training/Testing Procedures}
        Within subject pseudocategory decoding was then performed for each of the models on both datasets. A stratified k-fold cross-validation procedure was used to train and test the models. The data was stratified using the exemplar labels to ensure that the distribution of exemplars in each fold was representative of the distribution in the dataset. The number of folds was set to 12 for the Kaneshiro dataset\cite{kaneshiro_representational_2015} and 10 for the Gifford dataset\cite{gifford_large_2022}. The accuracy of the models was then calculated as the proportion of correctly classified trials. This resulted in 120 and 100 accuracy values for the Kaneshiro dataset\cite{kaneshiro_representational_2015} and Gifford datasets\cite{gifford_large_2022} respectively.

    \section{Statistical Analysis}
    To determine if the models were learning exemplar representations we compared the decoding accuracy of the models on the pseudocategory labels to chance accuracy using a one-sample t-test. The significance level was set at $\alpha=0.05$, and then Bonferroni correction was applied to account for multiple comparisons. Since we performed tests for 6 models over 2 datasets this resulted in the significance level being adjusted to $\alpha=\frac{0.05}{6\cdot2}\approx 0.0042$. 
    
    \begin{table}[b!]
        \centering
        \caption{Pseudocategory Decoding Accuracies}
        \label{tab:pseudocategory-decoding-accuracy}
        \begin{tabular}{lc}
            \hline
            \textbf{Model} & \textbf{Accuracy (\%)}\\ \hline
            \multicolumn{1}{l}{\textbf{Kaneshiro et al. (2015)}} & \\
            \quad kNN & $10.6809^*$ \\
            \quad SVM & $14.5370^*$ \\
            \quad LDA & $14.2477^*$  \\
            \quad EEGNet & $14.4483^*$ \\
            \quad DeepConvNet & $16.6532^*$ \\
            \quad ShallowConvNet & $12.8106^*$
            \\[1ex]
            \multicolumn{1}{l}{\textbf{Gifford et al. (2022)}} & \\
            \quad kNN & $22.1076^*$  \\
            \quad SVM & $30.1250^*$ \\
            \quad LDA & $30.3076^*$  \\
            \quad EEGNet & $30.1033^*$ \\
            \quad DeepConvNet & $30.5511^*$  \\
            \quad ShallowConvNet & $27.8163^*$ \\
            \hline
        \end{tabular}
    \end{table}

    \section{Results}

\begin{figure*}[t!]
        \centerline{\includegraphics[width=\textwidth]{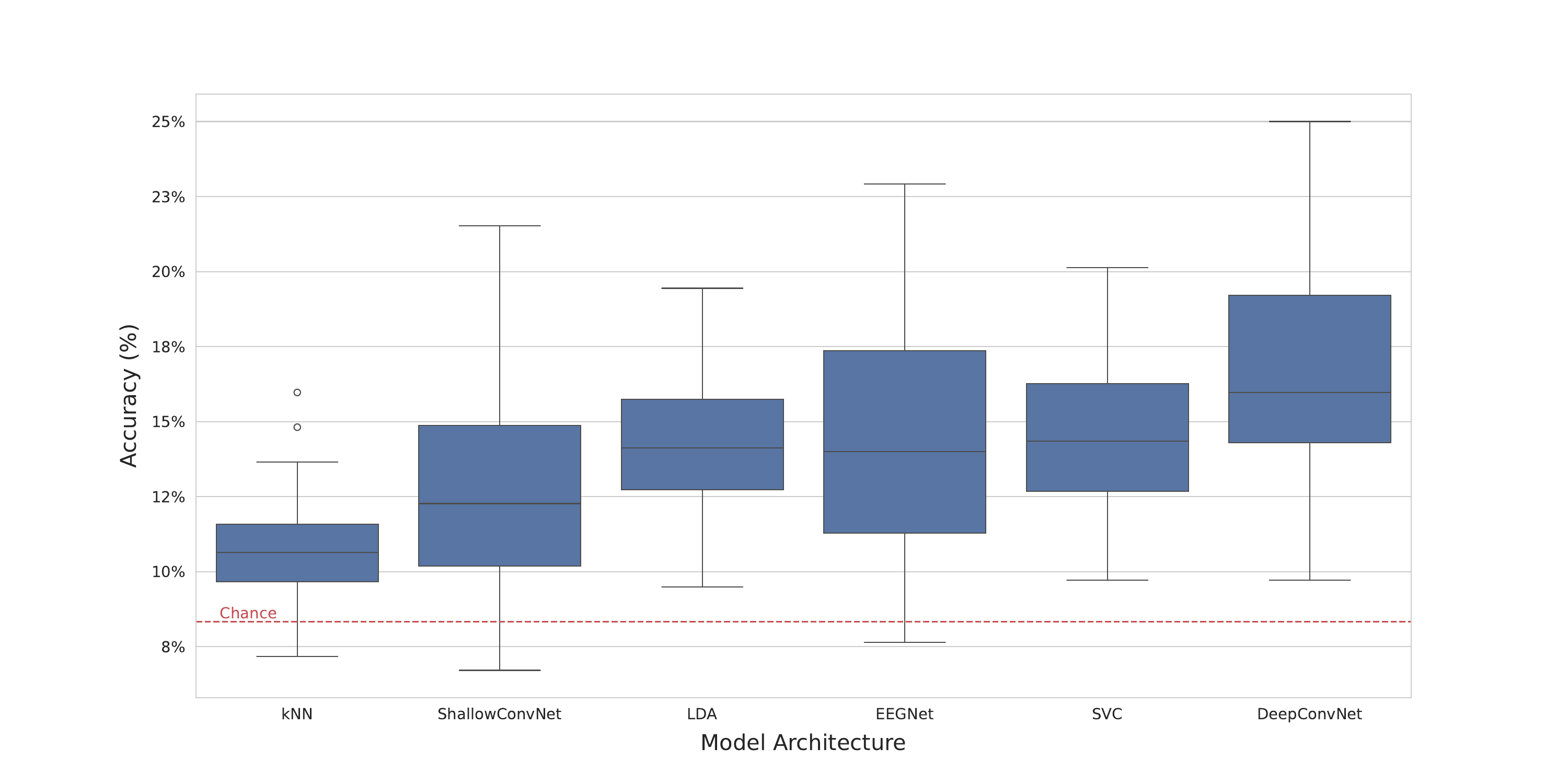}}
        \caption{A box plot of the pseudocategory decoding accuracies for each model trained using the Kaneshiro dataset\cite{kaneshiro_representational_2015}.}
        \label{fig:kaneshiro-accuracy}
        \end{figure*}

\begin{figure*}[t!]
        \centerline{\includegraphics[width=\textwidth]{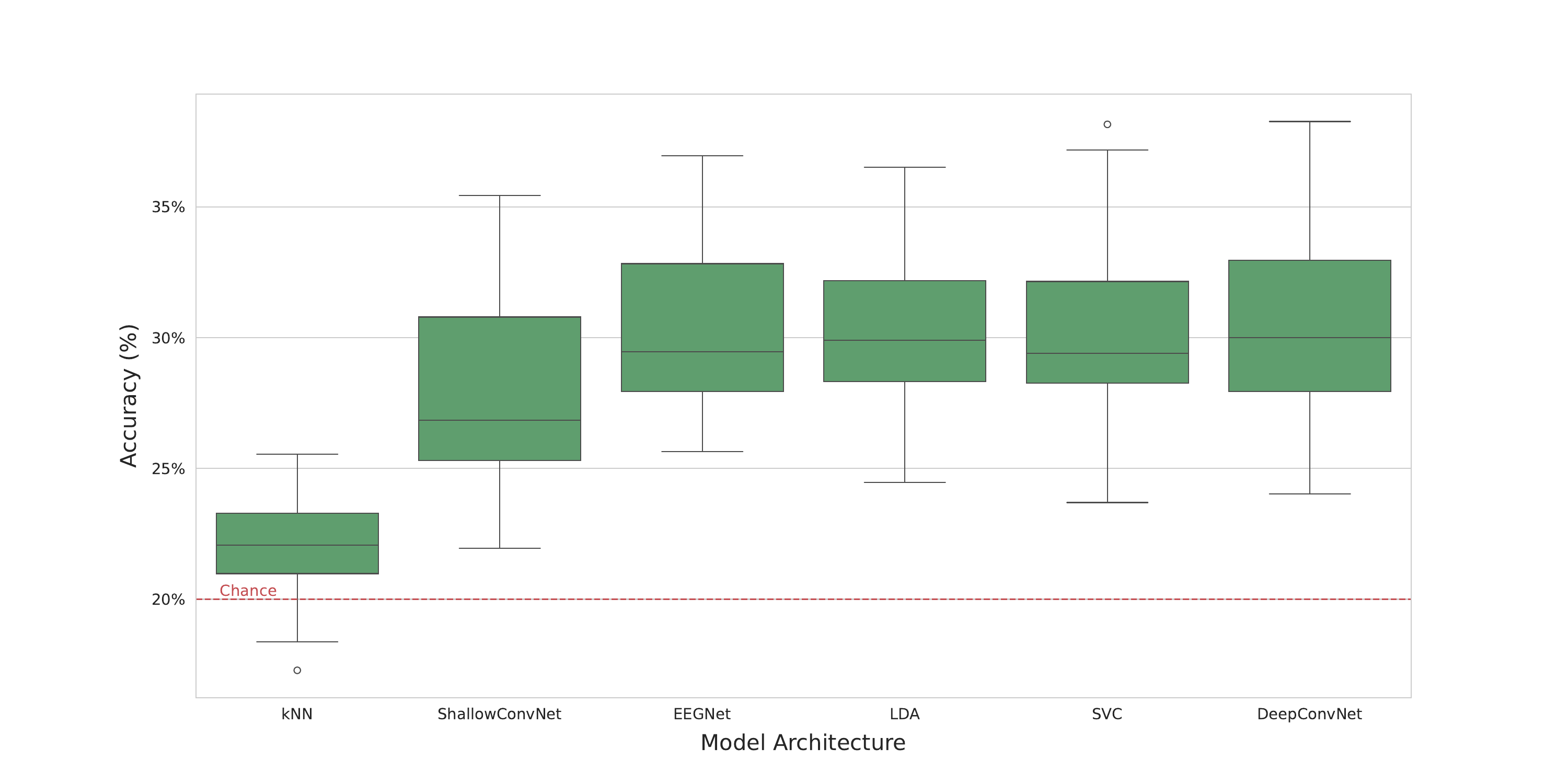}}
        \caption{A box plot of the pseudocategory decoding accuracies for each model trained using the Gifford dataset\cite{gifford_large_2022}.}
        \label{fig:gifford-accuracy}
        \end{figure*}

    The results of the statistical analysis are shown in Table~\ref{tab:pseudocategory-decoding-accuracy}. Starred values indicate that the accuracy was above chance with statistical significance (i.e. $p<0.001$, given a one-sample t-test with Bonferroni correction). Given that all models achieved decoding accuracies above chance with statistical significance on both datasets, our results show that classification algorithms are indeed able to learn exemplar representations from categories. While the effect may have been more dramatic for highly complex models, even models such as kNN with only a single parameter can exploit exemplar representations to some degree. Moreover, it is of particular interest that the DeepConvNet had an accuracy of over twice the chance accuracy on the Kaneshiro dataset\cite{kaneshiro_representational_2015}. This interesting because, as mentioned previously, the Kaneshiro dataset\cite{kaneshiro_representational_2015} has been used for category decoding in 12 additional studies following its publication; 10 of which used deep learning models. This suggests that the true accuracy of models affected by the leakage within the literature may differ dramatically from the published results. Box plots of the pseudocategory decoding accuracy are shown for both of the datasets in Fig.~\ref{fig:kaneshiro-accuracy} and Fig~\ref{fig:gifford-accuracy}.  Moreover, it is of interest that all the models performed above chance on the Gifford dataset\cite{gifford_large_2022}, even though it featured 23 exemplars per pseudocategory. This demonstrates that the classification algorithms are able to exploit exemplar representations even when the number of exemplars per category is much higher than in the Kaneshiro dataset\cite{kaneshiro_representational_2015}, and it is not a flaw which can be easily mitigated by increasing the number of stimuli used in experiments.
    \section{Future Work}
    While the results of this study demonstrate the ability of classification algorithms to learn exemplar representations given only category labels, further work is needed in recreating a sample of published models with reported decoding accuracies which are suspected to be inflated due to repeated exemplar leakage. This would allow a more comprehensive quantitative analysis of the impact of the leakage and clarify the true state-of-the-art in single-trial category decoding. Moreover, while we know that the models are capable of learning exemplar representations in a category decoding task, it has not yet been demonstrated that the same form of leakage can occur in other forms of EEG analysis. In particular if multiple categories are allowed to appear in a single image then instead of decoding a class we would be detecting the presence of classes. While it seems likely that the same form of leakage would occur in this case, it has not yet been demonstrated, though datasets featuring repeated stimuli and class presence labels are available\cite{xu_all_joined_2024}.

    \section{Conclusion}
    The results of this study demonstrate that given the multiple trials per stimulus, and improper model evaluation methods, almost any classification algorithm is capable of learning exemplar representations given only category labels. This has significant implications not only for the fields of neuroimaging and machine-learning, but also for any brain computer interface (BCI) applications which might be developed on the basis of results affected by such leakage. We achieved this using two datasets, the Kaneshiro dataset\cite{kaneshiro_representational_2015} and the Gifford dataset\cite{gifford_large_2022}, and a variety of prominent EEG classification algorithms. Use of the former dataset demonstrated that this leakage likely affects several published works. And use of the latter dataset established both the potential of this leakage to occur in other datasets, and that classification algorithms may still learn exemplar specific features, even when the number of exemplars per category is much higher than in the Kaneshiro dataset\cite{kaneshiro_representational_2015}. The significance of our findings is not only in highlighting the potential for leakage in single-trial category decoding studies which use repeated trials per stimulus, but also in revealing that the true state-of-the-art in single-trial category decoding is uncertain. It seems intuitive that as highest reported performances of models increase on this dataset, the extent to which they are likely affected by repeated exemplar leakage also increases. Therefore, which models have performed best at learning category representations is currently unclear.

\bibliographystyle{IEEEtran}
\bibliography{references}
\end{document}